\documentclass[english]{article}
\usepackage[T1]{fontenc}
\usepackage[latin9]{inputenc}
\usepackage{geometry}
\usepackage{amsmath}
\usepackage{amssymb}
\usepackage{esint}

\makeatletter

\providecommand{\tabularnewline}{\\}

\newcommand{\tmop}[1]{\ensuremath{\operatorname{#1}}}\newcommand{\tmtextbf}[1]{{\bfseries{#1}}}

\makeatother

\usepackage{babel}

\usepackage{cite}

\makeatother

\usepackage{babel}

\makeatother

\usepackage{babel}

\makeatother

\usepackage{babel}

\makeatother

\usepackage{babel}

\begin{document}
\begin{titlepage}

\begin{flushright}
LTH 902 
\par\end{flushright}

\vskip 1cm

\begin{center}
\textbf{\Large T-branes and Yukawa Couplings}\textbf{\large {} \vskip
1.2cm }{\large Chan-Chi Chiou}%
\footnote{e-mail: \texttt{chichi@liverpool.ac.uk}%
}{\large , Alon E. Faraggi}%
\footnote{e-mail: \texttt{faraggi@liverpool.ac.uk}%
}{\large {} , Radu Tatar}%
\footnote{e-mail: \texttt{rtatar@liverpool.ac.uk}%
}{\large {} and William Walters}%
\footnote{e-mail: \texttt{wrw23@liverpool.ac.uk}%
}{\large {} } 
\par\end{center}

\begin{center}
\vskip 0.7cm \textit{Division of Theoretical Physics, Department
of Mathematical Sciences, The University of Liverpool, Liverpool,
United Kingdom } \vskip 1.5cm \abstract{We consider various configurations
of T-branes which are non-abelian bound states of branes and were
recently introduced by Cecotti, Cordova, Heckman and Vafa. They are
a refinement of the concept of monodromic branes featured in phenomenological
F-theory models. We are particularly interested in the T-branes corresponding
to $Z_{3}$ and $Z_{4}$ monodromies, which are used to break $E_{7}$
or $E_{8}$ gauge groups to $SU(5)_{GUT}$. Our results imply that
the up-type and down-type Yukawa couplings for the breaking of $E_{7}$
are zero, whereas up-type and down-type Yukawa couplings, together
with right handed neutrino Yukawas are non-zero for the case of the
breaking of $E_{8}$. The dimension four proton decay mediating term
is avoided in models with either $E_{7}$ or $E_{8}$ breaking. } 
\par\end{center}

\vskip 5cm

January 2011 \end{titlepage}

{\tableofcontents{}}

\section{Introduction}

Recent studies have revealed that F-theory phenomenology can describe
realistic features of particle phenomenology \cite{dw1,bhv1,bhv2}
(see also \cite{tatar2,tatar3}). As compared to previous models for
particle phenomenology involving D5 branes (type IIB) or D6 branes
(type IIA), the F-theory approach localizes gauge fields on D7 branes
wrapped on four cycles, matter on complex curves inside the four cycles
and the Yukawa couplings at the intersection points of complex curves.
The main observation of \cite{dw1,bhv1,bhv2} was that all the fields
are described in an eight-dimensional topological field theory giving
rise to a four dimensional ${\it N}=1$ supersymmetric field theory.
The matter fields have Gaussian function on the directions normal
to the matter curves so they are genuinely trapped on such curves.
The F-theory trapping of matter curves can be nicely mapped into the
matter curve restrictions appearing in heterotic strings \cite{tatar2009}.

The formulation in terms of the eight-dimensional topological field
theory has led to a series of important developments which uncovered
various aspects of F-theory phenomenology (for recent reviews on the
subject see \cite{vafa,heckman,weigand}). One issue first pointed
out in \cite{tatar1} was related to the existence of novel features
like branch cuts in models giving rise to relevant Yukawa couplings.
This was due to the fact that, when studying such theories, the adjoint
field $\Phi$ describing the displacement orthogonal to the worldvolume
of the D7 branes was fixed to a background value $\left\langle \Phi\right\rangle $
taken to reside in the Cartan subalgebra. In order to achieve one
heavy mass generation, a new concept needed to be introduced which
is the seven brane monodromy. This means that, in order to deal with
F-theory compactifications on a Calabi-Yau with generic complex structure,
one needs to introduce branch cuts into the field theory and fields
have to be twisted by Weyl reflections at the branch cuts. To get
the Yukawa coupling in $SU(5)_{GUT}$, one needs to consider the breaking
$E_{6}\rightarrow SU(5)$ which involves a field $\Phi$ with the
above branch cuts.

Very recently an important step has been taken towards a better understanding
of this issue \cite{tbranes}. Instead of considering a diagonal $\left\langle \Phi\right\rangle $
with branch cuts, the authors of \cite{tbranes} have considered a
theory without branch cuts but with a non-diagonal $\left\langle \Phi\right\rangle $
and denoted such non-diagonal case as seven-branes {} {}``T-branes''.
Some example were worked out for the case of T-branes with $Z_{2}$
or $Z_{2}\times Z_{2}$ monodromy leading to breakings $E_{6}\rightarrow SU(5)$
or $E_{8}\rightarrow SU(5)$. The Yukawa couplings for the GUT group
were computed with the help of a residue formula.

The present work extends the results of \cite{tbranes} to the case
of $Z_{k}$ monodromy T-branes when $k=3$ and $k=4$. The $Z_{3}$
monodromy T-branes are used to break the $E_{7}$ group to $SU(5)_{GUT}\times SU\left(3\right)\times U(1)$
whereas the $Z_{4}$ monodromy T-branes are used to breaking the $E_{8}$
group to $SU(5)_{GUT}\times SU\left(4\right)\times U\left(1\right)$.
We consider the $E_{7}$ and $E_{8}$ breaking instead of just limiting
to the $E_{6}$ breaking for several reasons. From \cite{tatar2}
we know that we need the breaking of at least $E_{7}$ group in order
to get two $\bar{5}$ representations which would correspond to $\bar{5}_{M}$
and $\bar{5}_{H}$. This would allow a down-type Yukawa coupling but
prevent a proton decay term. The need for an $E_{7}$ gauge theory
was also considered in \cite{taizanapril2010} in order to avoid mixing
between the resolution cycles.

On the other hand, we also know from \cite{vafa2} that the important
feature of massive right handed neutrinos cannot be imposed at an
$E_{7}$ point of enhancement. Our computations in the case of the
breaking $E_{7}\rightarrow SU(5)_{GUT}$, by using the residue results,
show that both up-type and down-type Yukawa are zero for this case.
We then proceed to the $Z_{4}$ monodromy model breaking $E_{8}$
to $SU(5)_{GUT}$. In this case all the matter curves have singularity
at the origin but the residue results give non-zero Yukawa couplings,
including for the Dirac neutrino mass term.

Another important issue for any phenomenological model is proton longevity.
In addition to the matter--matter--Higgs couplings that are needed
to generate the fermion mass spectrum, supersymmetric theories give
rise to matter--matter--matter couplings that may yield dimension
four and five baryon and lepton number violating operators. This is
an intricate problem in Grand Unified Theories in general, and in
string theories in particular. The reason being that while proton
decay mediating operators must be adequately suppressed, Majorana
neutrino masses require lepton number violation. These two key phenomenological
constraints can be accommodated simultaneously, either by allowing
lepton number, while forbidding baryon number, violating operators,
or by using Dirac mass terms to generate the left--handed neutrino
masses. However, in Grand Unified Theories that admit $SO(10)$ embedding
of the Standard Model matter states only the former is possible due
to lepton--quark mass relations that are dictated by the larger gauge
symmetry. An appealing proposition to resolve this conundrum is the
existence of gauged $U(1)$ symmetries that forbid the baryon number
violating operators, while allowing lepton number violation. An example
of a well know symmetry that partially does the job is that of gauged
$U(1)_{B-L}$, which resides inside the GUT $SO(10)$. However, while
$U(1)_{B-L}$ does forbid the dimension four lepton and baryon number
violating operators, it does not forbid such dimension five operators,
and is therefore not sufficient. Furthermore, $U(1)_{B-L}$ forbid
the Majorana mass terms that are needed to generate light--neutrino
masses. Hence, the desired symmetries must extend the $U(1)_{B-L}$
symmetry and reside outside $SO(10)$. In perturbative heterotic string
theories the caveatted symmetries arise from the three generators
in the Cartan sub--algebra of the observable $E_{8}$, and the existence
of adequate phenomenological combinations has been explored in the
literature \cite{Faraggi:1994cv,Faraggi:2000cm,Pati:1996fn,Coriano:2007ba}.
In the present work we discuss the absence of proton decay within
heterotic F--theory constructions, which naturally contain $U(1)$
symmetries residing outside $SO(10)$.

We start in section 2 with a brief discussion of the concept of monodromic
branes introduced in \cite{tatar1} and the concept of T-branes introduced
in \cite{tbranes}. We then continue with presenting the case of $Z_{3},SU(3)$
and $Z_{4},SU(4)$ backgrounds. Section 3 is the main section of this
work, where we derive in detail the Yukawa couplings for the charged
particles and for the singlets in the case of the $Z_{3}$ breaking
of $E_{7}$ to $SU(5)_{GUT}$ and $Z_{4}$ breaking of $E_{8}$ to
$SU(5)_{GUT}$. The details of the section 2 and section 3 computations
are relegated to the Appendix where we also include some general $Z_{n}$
results.

\section{T-branes}

\subsection{Monodromic Branes}

The usual way to deal with field theory on D-branes is to identify
fields in the adjoint representations of gauge groups on branes with
directions orthogonal to the D-branes. This has been used extensively
for D3 branes probing singularities and D5 branes wrapped on resolution
2-cycles. The same method has been initiated for the case of D7 branes
wrapped on 4-cycles for intersection of two D7-branes in \cite{kv1},
based on the findings of \cite{km}.

The results of \cite{dw1,bhv1,bhv2} allow a further exploration of
the results of \cite{km}. Considering a four dimensional cycle $S$
with complex coordinates $u_{m},~m=1,2$, the effective theory of
zero modes along S can be described by an 8-dimensional field theory
with four directions along S whose content is given in Table 1 where
all the fields have their values in Lie algebra determined by the
singularity along S.

%
\begin{table}[t]
 \centering{}\begin{tabular}{l|c|cc|cc}
 & vector  & chiral  & multiplet  & anti-chiral  & multiplet \tabularnewline
\hline 
Bosonic fields  & $A_{\mu}$  & $\Phi_{mn}du_{m}\wedge du_{n}$  & $A_{\bar{m}}d\bar{u}_{\bar{m}}$  & $\overline{\Phi}_{\bar{m}\bar{n}}d\bar{u}_{\bar{m}}\wedge d\bar{u}_{\bar{n}}$  & $A_{m}du_{m}$ \tabularnewline
Fermionic fields  & $\eta$  & $\chi_{mn}du_{m}\wedge du_{n}$  & $\psi_{\bar{m}}d\bar{u}_{\bar{m}}$  & $\bar{\chi}_{\bar{m}\bar{n}}d\bar{u}_{\bar{m}}\wedge d\bar{u}_{\bar{n}}$  & $\bar{\psi}_{m}du_{m}$ \tabularnewline
\end{tabular}\caption{\label{tab:BHV-fields} Field contents on $S$. }
\end{table}


The field $\overline{\Phi}_{\bar{m}\bar{n}}d\bar{u}_{\bar{m}}\wedge d\bar{u}_{\bar{n}}$
represents the transverse fluctuations of the D7 branes in the Calabi-Yau
compactifications. The local geometry of the F-theory compactification
can describe deformation of singularities which, in turn, can be related
to describe matter and Yukawa couplings. In terms of the compactification
of F-theory on Calabi-Yau 4-folds, the vacuum expectation value for
the field $\left\langle \Phi\right\rangle $ corresponds to local
geometries of Calabi-Yau 4-fold and a mapping between the Calabi-Yau
geometries and the values of $\left\langle \Phi\right\rangle $ has
been used first in \cite{kv1}. The approach involves a parametrization
of a generic deformation of ADE singularities by $\mathfrak{h}\otimes C/W$
where $\mathfrak{h}$ is the Cartan subalgebra of the ADE algebra
and $W$ the Weyl group. On the other hand, for local Calabi--Yau
geometry that are fibrations of deformation of a singularity over
base space $B$, the local geometry is described using the above field
theory with the background values for $\Phi$ lying in $\mathfrak{h}\otimes C$
and varying over $B$. The generic deformations can be easily mapped
into the field theory quantities \cite{km} if one ignores the difference
between $\mathfrak{h}\otimes C/W$ and $\mathfrak{h}\otimes C$.

In the case of a non-compact complex curve $B$, the gauge group $\mathfrak{g}$
has one-rank larger than the singularity that remains over $B$ after
the deformation. The Weyl group is not important in this case and
mapping is clear \cite{kv1}. A more complicated case appears when
$B$ is a complex surface and the rank of the singularity group decreases
by two.

The Weyl group becomes important and the mapping becomes less clear
if the gauge group $\mathfrak{g}$ has a rank which is larger by at
least two than the singularity group that remains over $B$ after
the deformation. The identification of \cite{km} was extended in
\cite{tatar1} to the case when generic deformations decrease the
rank by two and the result was that the field $\left\langle \Phi\right\rangle $
had an unwanted feature of having branch cuts. As a simple example
discussed in that paper, let us consider the deformation of the singularity
$A_{N+1}\rightarrow A_{N-1}$ given by two parameters $s_{1}$ and
$s_{2}$ : \begin{equation}
Y^{2}=X^{2}+Z^{N}(Z^{2}+s_{1}Z+s_{2}),\end{equation}
 and we consider the identification between the two parameters $s_{1}$
and $s_{2}$ (which are related to the non-zero values of $\left\langle \Phi\right\rangle $)
and the coordinates of the complex surface $(u_{1},u_{2})$ as \begin{equation}
s_{1}=2u_{1},~~~s_{2}=u_{2},\end{equation}
 the non-zero part of $\left\langle \Phi\right\rangle $ has the values
$(\sqrt{u_{1}^{2}-u_{2}},-\sqrt{u_{1}^{2}-u_{2}})$ on the diagonal
which acquire a minus sign around the branch locus $u_{1}^{2}-u_{2}=0$.

One can try to generalize this model to the case with $A_{N+2}\rightarrow A_{N-1}$
given by two parameters $s_{1}$, $s_{2}$, $s_{3}$: \begin{equation}
Y^{2}=X^{2}+Z^{N}(Z^{3}+s_{1}Z^{2}+s_{2}Z+s_{3}),\label{Z33}\end{equation}
 where $s_{1}$, $s_{2}$, $s_{3}$ would be directly related to the
three non-zero diagonal entries for $\left\langle \Phi\right\rangle $.
There is no isomorphism relation between $s_{i}$ and $u_{i}$ so
a corresponding $Z_{3}$ model cannot be used to describe intersecting
branes.

This effect showed that the seven-brane monodromy was a required ingredient
in describing F-theory phenomenology and was subsequently developed
along other directions \cite{bouchard,vafa2,mssjune09,taizanoct09,dudas}.

\subsection{T-branes}

The use of monodromic branes was based on the assumption that the
field $\Phi$ is valued in the Cartan subalgebra. Very recently, the
work of \cite{tbranes} used models where $\left\langle \Phi\right\rangle $
is upper triangular on some locus and such a configuration of seven-branes
was denoted as T-branes, without the unwanted branch cuts on the field
theory side. The difference between the T-branes and the intersecting
branes lies in dealing with the spectral equation \begin{equation}
P_{\Phi}(z)=\det(z-\Phi)=0\end{equation}
 When $\Phi$ belongs to the Cartan subalgebra, the spectral equation
becomes \begin{equation}
\prod_{i}(z-\lambda_{i})=0\end{equation}
 where $\lambda_{i}$ are the eigenvalues of $\Phi$ and they denote
the directions of the intersecting branes. In case of non-diagonalizable
Higgs fields, the spectral equation does not have a geometric interpretation
and the intersecting branes picture does not hold, the monodromy group
is now encoded in the form of the spectral equation. The paper \cite{tbranes}
showed that the branch cuts are removed for the $Z_{2}$ model. One
interesting aspect is that the T-brane model can consider more complicated
$Z_{n},n>2$ cases like (\ref{Z33}) which we now describe.

\subsection{$SU\left(3\right)$}

Let us consider the spectral equation for an $SU(3)$ field: \begin{equation}
P_{\Phi}(z)=z^{3}-x\end{equation}
 for which there is a $Z_{3}$ monodromy. In the holomorphic gauge
the Higgs field is: \begin{equation}
\Phi=\left(\begin{array}{ccc}
0 & 1 & 0\\
0 & 0 & 1\\
x & 0 & 0\end{array}\right).\end{equation}
 which is an intermediate case between a diagonal background and a
nilpotent Higgs field \begin{equation}
\Phi=\left(\begin{array}{ccc}
0 & 1 & 0\\
0 & 0 & 1\\
0 & 0 & 0\end{array}\right).\end{equation}
 We transform this to unitary gauge by using a positive diagonal matrix
$g$ with unit determinant: \begin{equation}
g=\left(\begin{array}{ccc}
e^{f_{1}} & 0 & 0\\
0 & e^{f_{2}} & 0\\
0 & 0 & e^{f_{3}}\end{array}\right)\end{equation}
 with the condition that $f_{1}+f_{2}+f_{3}=0$, and the $f_{a}$
are real. Solving the D-term equation \begin{equation}
\omega\wedge F_{A}+\frac{i}{2}\left[\Phi^{\dagger},\Phi\right]=0\end{equation}
 should give us the Toda equation \begin{equation}
\Delta f_{a}=C_{ab}e^{f_{b}},\end{equation}
 where $C_{ab}$ is the Cartan matrix of $SU\left(3\right)$ which
is given by \begin{equation}
C_{ab}=\left(\begin{array}{cc}
2 & -1\\
-1 & 2\end{array}\right).\end{equation}
 This presents an apparent problem, since there are three $f_{a}$'s
but $C_{ab}$ is merely a 2x2 matrix. However this is because the
unit determinant requirement of $g$ means that there are only two
linearly independent $f_{a}$'s so the equation makes sense. We only
end up with the required Toda equation if we take the three $f_{a}$'s
in $g$ as specific linear combinations of the two linearly independent
functions which we call $h_{a}$ in Appendix.

As derived in the Appendix, the components for the unitary transformation
for the nilpotent field $\Phi$ satisfy \begin{eqnarray}
\partial\overline{\partial}f_{1} & = & 2e^{f_{1}}-e^{f_{2}}\nonumber \\
\partial\overline{\partial}f_{2} & = & -e^{f_{1}}+2e^{f_{2}}\end{eqnarray}
 and, for the general case \begin{equation}
\Phi=\left(\begin{array}{ccc}
0 & 1 & 0\\
0 & 0 & 1\\
x & 0 & 0\end{array}\right),\end{equation}
 we get, for $f_{a}$ depending only on $r$, the following two equations:
\begin{eqnarray}
\left(\frac{d^{2}}{dr^{2}}+\frac{1}{r}\frac{d}{dr}\right)f_{1} & = & r^{\frac{2}{3}}\left(2e^{f_{1}}-e^{-f_{1}-f_{2}}-e^{f_{2}}\right)\nonumber \\
\left(\frac{d^{2}}{dr^{2}}+\frac{1}{r}\frac{d}{dr}\right)f_{2} & = & r^{\frac{2}{3}}\left(2e^{f_{2}}-e^{-f_{1}-f_{2}}-e^{f_{1}}\right).\end{eqnarray}

This set of equations generalizes the D-term equation \begin{equation}
\left(\frac{d^{2}}{ds^{2}}+\frac{1}{s}\frac{d}{ds}\right)=\frac{1}{2}\mbox{sinh}(2f),~~s=\frac{8}{3}r^{3/2}\end{equation}
 obtained in \cite{tbranes} for the $Z_{2}$ T-branes which was a
special instance of the Painleve III differential equation whose asymptotic
behaviour was nicely mapped into the diagonalizable intersecting brane
case for $r\rightarrow\infty$ and a nilpotent Higgs for $r\rightarrow0$.

The asymptotic regions are represented by the case when $r$ is either
very small or very large. In our case, for $r$ very small, the condition
on $g$ to be everywhere non-singular implies that near $r=0$ the
functions $f_{1},f_{2}$ have logarithmic singularities and their
exponentials approach non-zero constant matrices in the Cartan $U(1)^{2}$
of $SU(3)$. On the other hand, for large values of $r$ we expect
to get the case of intersecting branes obtained when the value of
the flux $F_{A}$ is zero. An explicit solutions for the $f_{1},f_{2}$
should obey both limits for small and large $r$. We expect that a
physically valid solution to exist and the configuration to be supersymmetric
but a full solution involves generalizing the solution of Painleve
III differential equation to the case of 2 functions.

\subsubsection{Brane recombination}

We consider infinitesimal perturbations to the holomorphic Higgs field
of the form: \begin{equation}
\varphi=\mathrm{\tmop{ad}}_{\Phi}\left(\xi\right)+h\end{equation}
 where $\xi$ is an arbitrary gauge transformation. Start with a $U\left(3\right)$
gauge theory, which can be thought of as corresponding to three superimposed
D7-branes, and deform this theory using the $SU\left(3\right)$ Higgs
vev \begin{equation}
\Phi=\left(\begin{array}{ccc}
0 & 1 & 0\\
0 & 0 & 1\\
x & 0 & 0\end{array}\right).\end{equation}
 Consider the action of this field on an arbitrary gauge field \begin{equation}
\xi=\left(\begin{array}{ccc}
a & b & c\\
d & e & f\\
g & h & i\end{array}\right),\end{equation}
 as \begin{equation}
\mathrm{\tmop{ad}}_{\Phi}\left(\xi\right)=\left[\Phi,\xi\right]=\left(\begin{array}{ccc}
d-cx & e-a & f-b\\
g-fx & h-d & i-e\\
ax-ix & bx-g & cx-h\end{array}\right).\end{equation}
 We then see that we can set certain components of the field $h$
to zero. Firstly note that $\mathrm{\tmop{ad}}_{\Phi}\left(\xi\right)$
is traceless so we can remove two diagonal degrees of freedom from
$h$. Then we can set $h_{12}$ and $h_{23}$ to zero using $e-a$
and $i-e$ respectively. However, after doing this, we do not have
any freedom to set $h_{31}$ to zero since \begin{equation}
ax-ix=-x(e-a)-x(i-e)\end{equation}
 Similarly we can set $h_{13}$ and $h_{21}$ to zero but then this
fixes the gauge transform that can be made on $h_{32}$ since \begin{equation}
bx-g=-x\left(f-b\right)-\left(g-fx\right).\end{equation}
 So the most general perturbation that can be made after gauge fixing
is \begin{equation}
\varphi=\left(\begin{array}{ccc}
\frac{1}{3}\alpha\left(x,y\right) & 0 & 0\\
0 & \frac{1}{3}\alpha\left(x,y\right) & 0\\
\gamma\left(x,y\right) & \beta\left(x,y\right) & \frac{1}{3}\alpha\left(x,y\right)\end{array}\right).\end{equation}
 With this perturbation, the spectral equation is deformed as \begin{equation}
z^{3}-x\rightarrow\left(z-\frac{1}{3}\alpha\left(x,y\right)\right)\left(\left(z-\frac{1}{3}\alpha\left(x,y\right)\right)^{2}-\beta\left(x,y\right)\right)-\left(x+\gamma\left(x,y\right)\right)\end{equation}
 which, to first order in the perturbation, is \begin{equation}
z^{3}-z^{2}\alpha\left(x,y\right)-z\beta\left(x,y\right)-x-\gamma\left(x,y\right).\end{equation}
 After changing coordinates to \begin{equation}
\left(\tilde{x},\tilde{y},\tilde{z}\right)=\left(z,y,P_{\Phi}\left(z\right)\right),\end{equation}
 this becomes, in terms of the new brane worldvolume $\tilde{z}=0$,
\begin{equation}
\tilde{z}-\left(\tilde{x}^{2}\alpha\left(\tilde{x}^{3},\tilde{y}\right)+\tilde{x}\beta\left(\tilde{x}^{3},\tilde{y}\right)+\gamma\left(\tilde{x}^{3},\tilde{y}\right)\right).\end{equation}
 Hence the perturbations $\alpha$, $\beta$ and $\gamma$ just make
up the components of order $\tilde{x}^{3n+2}$, $\tilde{x}^{3n+1}$
and $\tilde{x}^{3n}$ of the Taylor expansion in $\tilde{x}$ of a
single $U\left(1\right)$ field. This is interpreted as the three
D7-branes recombining into a single D7-brane.

The Kahler metric on this brane can be determined by the pullback
of the flat Kahler metric onto the brane. We start from the flat Kahler
metric \begin{equation}
\omega=\frac{i}{2}\left(dx\wedge d\overline{x}+dy\wedge d\overline{y}+dz\wedge d\overline{z}\right),\end{equation}
 change to the new coordinates, and note that on the brane we have
$\tilde{z}=P_{\Phi}\left(z\right)=z^{3}-x=0$, so that $x=z^{3}=\tilde{x}^{3}$,
we then have \begin{eqnarray}
dx & = & 3\tilde{x}^{2}d\tilde{x}\nonumber \\
dy & = & d\tilde{y}\nonumber \\
dz & = & d\tilde{x}.\end{eqnarray}
 The Kahler form is \begin{equation}
\omega=\frac{i}{2}\left(\left(1+9\left|\tilde{x}\right|^{4}\right)d\tilde{x}\wedge d\overline{\tilde{x}}+d\tilde{y}\wedge d\overline{\tilde{y}}\right),\end{equation}
 so the recombined brane is curved, as in the $SU\left(2\right)$
case \cite{tbranes}.

\subsection{$SU\left(4\right)$}

Let us now consider the spectral equation for an $SU(4)$ field: \begin{equation}
P_{\phi}(z)=z^{4}-x,\end{equation}
 for which there is a $Z_{4}$ monodromy. In the holomorphic gauge
the Higgs field becomes \begin{equation}
\Phi=\left(\begin{array}{cccc}
0 & 1 & 0 & 0\\
0 & 0 & 1 & 0\\
0 & 0 & 0 & 1\\
x & 0 & 0 & 0\end{array}\right).\end{equation}
 which is an intermediate between a diagonal Higgs field and a nilpotent
Higgs field in holomorphic gauge \begin{equation}
\Phi=\left(\begin{array}{cccc}
0 & 1 & 0 & 0\\
0 & 0 & 1 & 0\\
0 & 0 & 0 & 1\\
0 & 0 & 0 & 0\end{array}\right),\end{equation}
 As derived explicitly in the Appendix, we require that the components
of the unitary transformation satisfy. \begin{eqnarray}
\partial\overline{\partial}f_{1} & = & 2e^{f_{1}}-e^{f_{2}}\nonumber \\
\partial\overline{\partial}f_{2} & = & -e^{f_{1}}+2e^{f_{2}}-e^{f_{3}}\nonumber \\
\partial\overline{\partial}f_{3} & = & -e^{f_{2}}+2e^{f_{3}}.\end{eqnarray}
 which have the desired form \begin{equation}
\partial\overline{\partial}f_{a}=C_{ab}e^{f_{b}},\end{equation}
 where $C_{ab}$ is the $SU\left(4\right)$ Cartan matrix \begin{equation}
\left(\begin{array}{ccc}
2 & -1 & 0\\
-1 & 2 & -1\\
0 & -1 & 2\end{array}\right).\end{equation}

For the general case, we proceed the same way as with $SU\left(3\right)$
(as derived in the Appendix) and get the equations for $f_{i}$ as
\begin{eqnarray}
\partial\overline{\partial}f_{1} & = & r^{\frac{1}{2}}\left(2e^{f_{1}}-e^{f_{2}}-e^{-f_{1}-f_{2}-f_{3}}\right)\nonumber \\
\partial\overline{\partial}f_{2} & = & r^{\frac{1}{2}}\left(-e^{f_{1}}+2e^{f_{2}}-e^{f_{3}}\right)\nonumber \\
\partial\overline{\partial}f_{3} & = & r^{\frac{1}{2}}\left(-e^{f_{2}}+2e^{f_{3}}-e^{-f_{1}-f_{2}-f_{3}}\right).\end{eqnarray}
 This set of equations is again similar to the one obtained in \cite{tbranes}
for the $Z_{2}$ T-branes and we are unaware of any known solution
for the set of differential equation for $f_{1},f_{2},f_{3}$.

\subsubsection{Brane recombination}

As was done in the $SU\left(3\right)$ case, we consider infinitesimal
perturbations to the Higgs field and then see which can be gauged
away to zero by a $U\left(4\right)$ gauge transformation. The result
here turns out be that the most general perturbation after gauge fixing
is \begin{equation}
\varphi=\left(\begin{array}{cccc}
\frac{1}{4}\alpha\left(x,y\right) & 0 & 0 & 0\\
0 & \frac{1}{4}\alpha\left(x,y\right) & 0 & 0\\
0 & 0 & \frac{1}{4}\alpha\left(x,y\right) & 0\\
\delta\left(x,y\right) & \gamma\left(x,y\right) & \beta\left(x,y\right) & \frac{1}{4}\alpha\left(x,y\right)\end{array}\right),\end{equation}
 which means that the spectral equation is now \begin{equation}
\left(z-\frac{1}{4}\alpha\left(x,y\right)\right)\left(\left(z-\frac{1}{4}\alpha\left(x,y\right)\right)\left(\left(z-\frac{1}{4}\alpha\left(x,y\right)\right)^{2}-\beta\left(x,y\right)\right)-\gamma\left(x,y\right)\right)-\left(x+\delta\left(x,y\right)\right),\end{equation}
 expanding this to first order in the perturbation, one obtains \begin{equation}
z^{4}-x-z^{3}\alpha\left(x,y\right)-z^{2}\beta\left(x,y\right)-z\gamma\left(x,y\right)-\delta\left(x,y\right).\end{equation}
 Then, changing coordinates to \begin{equation}
\left(\tilde{x},\tilde{y},\tilde{z}\right)=\left(z,y,P_{\Phi}\left(z\right)\right),\end{equation}
 (where here $P_{\Phi}\left(z\right)=z^{4}-x$ is the original spectral
equation before the perturbations) this becomes\tmtextbf{} \begin{equation}
\tilde{z}-\left(\tilde{x}^{3}\alpha\left(\tilde{x}^{4},\tilde{y}\right)-\tilde{x}^{2}\beta\left(\tilde{x}^{4},\tilde{y}\right)-\tilde{x}\gamma\left(\tilde{x}^{4},\tilde{y}\right)-\delta\left(\tilde{x}^{4},\tilde{y}\right)\right).\end{equation}
 So, as with the $SU\left(3\right)$ case, the seemingly distinct
fields are actually just components of a single field, so the effect
of the Higgs vev is to recombine the four superimposed D7-branes with
a $U\left(4\right)$ gauge group into a single D7-brane with a $U\left(1\right)$
gauge group.

Similarly to the $SU\left(4\right)$ case, the Kahler form on this
recombined brane is given by \begin{equation}
\omega=\frac{i}{2}\left(\left(1+16\left|\tilde{x}\right|^{6}\right)d\tilde{x}\wedge d\overline{\tilde{x}}+d\tilde{y}\wedge d\overline{\tilde{y}}\right),\end{equation}
 so this recombined brane is also curved.

\section{GUT Models and $Z_{k},k=3,4$ monodromy}

This is the main section of our work and we use the backgrounds of
the previous section together with the approach originated in \cite{tbranes}
to derive formulas for various types of Yukawa couplings for $SU(5)$
F-theory GUT.

We are going to break either $E_{7}$ with an $Z_{3}$ T-brane and
or $E_{8}$ with a $Z_{4}$ T-brane. Our conclusion is that the $E_{7}$
breaking gives rise to null up-type and down-type Yukawa couplings
whereas the $E_{8}$ model gives rise to non-zero Yukawa couplings
and also removes the proton decay term. We are also considering the
singlet couplings (right handed neutrinos) and show that they are
non-zero for the $E_{8}$ breaking. The Majorana masses are not allowed
as we need the $U(1)$ symmetries.

\subsection{Computation of Yukawa couplings for Localized modes}

\subsubsection{$E_{7}\rightarrow SU\left(5\right)\times SU\left(3\right)\times U\left(1\right)$}

Here we use an $SU\left(3\right)\times U\left(1\right)$ Higgs field
which preserves an unbroken $SU\left(5\right)$: \begin{equation}
\Phi=\left(\begin{array}{ccc}
0 & 1 & 0\\
0 & 0 & 1\\
x & 0 & 0\end{array}\right)\oplus\left(y\right).\end{equation}
 The adjoint of $E_{7}$ decomposes under this breaking as \begin{equation}
\mathbf{133}\rightarrow\left(\mathbf{1},\mathbf{1}\right)_{0}\oplus\left(\mathbf{1},\mathbf{8}\right)_{0}\oplus\left(\mathbf{24},\mathbf{1}\right)_{0}\oplus\left(\mathbf{\overline{5}},\mathbf{3}\right)_{-2}\oplus\left(\mathbf{5},\mathbf{\overline{3}}\right)_{2}\oplus\left(\mathbf{\overline{10}},\mathbf{3}\right)_{1}\oplus\left(\mathbf{10},\mathbf{\overline{3}}\right)_{-1}\oplus\left(\mathbf{5},\mathbf{1}\right)_{-3}\oplus\left(\mathbf{\overline{5}},\mathbf{1}\right)_{3}.\end{equation}
 By noting that the required interaction terms are of the form \begin{equation}
\mathbf{5}_{H}\cdot\mathbf{10}_{M}\cdot\mathbf{10}_{M}\hspace{0.25em}\hspace{0.25em}\hspace{0.25em}\hspace{0.25em}\hspace{0.25em}\mathrm{\tmop{and}}\hspace{0.25em}\hspace{0.25em}\hspace{0.25em}\hspace{0.25em}\hspace{0.25em}\mathbf{\overline{5}}_{H}\cdot\mathbf{\overline{5}}_{M}\cdot\mathbf{10}_{M},\end{equation}
 and looking at the $U\left(1\right)$ charges in the decomposition,
we see that we can identify the $\left(\mathbf{5},\mathbf{\overline{3}}\right)_{2}$
and the $\left(\mathbf{\overline{5}},\mathbf{3}\right)_{-2}$ as the
$\mathbf{5}_{H}$ and $\mathbf{\overline{5}}_{H}$, and the $\left(\mathbf{\overline{5}},\mathbf{1}\right)_{3}$
as the $\mathbf{\overline{5}}_{M}$.

Looking at the $\mathbf{\overline{5}}_{M}$, we see that under the
action of $\Phi$ the mode is simply multiplied by $3y$, which is
obviously only invertible away from $y=0$ and so the torsion equation
is solved with a matter curve $y=0$ and we have \begin{equation}
\eta_{\mathbf{\overline{5}}_{M}}=\frac{1}{3}\varphi_{\mathbf{\overline{5}}_{M}}.\end{equation}
 The $\mathbf{10}_{M}$ is in the antifundamental of the $SU\left(3\right)$
\begin{equation}
\varphi_{\mathbf{10}_{M}}=\left(\begin{array}{c}
\varphi_{\mathbf{10}_{M}}^{1}\\
\varphi_{\mathbf{10}_{M}}^{2}\\
\varphi_{\mathbf{10}_{M}}^{3}\end{array}\right),\end{equation}
 To see how this transforms under the Higgs field, we use the following
basis: \begin{equation}
\left(\begin{array}{c}
e_{2}\wedge e_{3}\\
e_{3}\wedge e_{1}\\
e_{1}\wedge e_{2}\end{array}\right),\end{equation}
 where the $e_{a}$ span the fundamental of $SU\left(3\right)$ and
the matrix acts on each component as \begin{equation}
\Phi\left(e_{a}\wedge e_{b}\right)=\left(\Phi e_{a}\right)\wedge e_{b}+e_{a}\wedge\left(\Phi e_{b}\right).\end{equation}
 So under a gauge transformation we have \begin{equation}
\delta\varphi_{\mathbf{10}_{M}}=\left(\begin{array}{ccc}
-2y & 0 & -x\\
-1 & -2y & 0\\
0 & -1 & -2y\end{array}\right)\left(\begin{array}{c}
a\\
b\\
c\end{array}\right).\end{equation}
 Since this matrix is invertible when the determinant is non-zero,
i.e. away from $-8y^{3}-x=0$, then $\varphi_{\mathbf{10}_{M}}$ is
gauge equivalent to zero away from this locus, so clearly the matter
curve is defined by \begin{equation}
f=-8y^{3}-x.\end{equation}
 Notice that this matter curve is self-intersecting, as is required
in order to have the coupling $\mathbf{5}_{H}\cdot\mathbf{10}_{M}\cdot\mathbf{10}_{M}$.
On the matter curve, we can still set to zero the last two components
of $\varphi_{\mathbf{10}_{M}}$, so we have \begin{equation}
\varphi_{\mathbf{10}_{M}}=\left(\begin{array}{c}
\varphi_{\mathbf{10}_{M}}^{1}\\
0\\
0\end{array}\right).\end{equation}
 The torsion equation can be solved using the adjugate matrix: \begin{equation}
\eta_{\mathbf{10}_{M}}=\left(\begin{array}{ccc}
4y^{2} & x & -2xy\\
-2y & 4y^{2} & x\\
1 & -2y & 4y^{2}\end{array}\right)\left(\begin{array}{c}
\varphi_{\mathbf{10}_{M}}^{1}\\
0\\
0\end{array}\right)=\left(\begin{array}{c}
4y^{2}\varphi_{\mathbf{10}_{M}}^{1}\\
-2y\varphi_{\mathbf{10}_{M}}^{1}\\
\varphi_{\mathbf{10}_{M}}^{1}\end{array}\right).\end{equation}
 The $\mathbf{5}_{H}$ transforms in the $\mathbf{\overline{3}}$
of $SU\left(3\right)$ and so we can use the result of the $\mathbf{10}_{M}$,
just replacing $y$ with $-2y$ because of the different $U\left(1\right)$
charge so the solution to the torsion equation is \begin{equation}
\eta_{\mathbf{5}_{H}}=\left(\begin{array}{ccc}
16y^{2} & x & 4xy\\
4y & 16y^{2} & x\\
1 & 4y & 16y^{2}\end{array}\right)\left(\begin{array}{c}
\varphi_{\mathbf{5}_{H}}^{1}\\
0\\
0\end{array}\right)=\left(\begin{array}{c}
16y^{2}\varphi_{\mathbf{5}_{H}}^{1}\\
4y\varphi_{\mathbf{5}_{H}}^{1}\\
\varphi_{\mathbf{5}_{H}}^{1}\end{array}\right),\end{equation}
 and the matter curve is given by $f=64y^{3}-x$.

The $\mathbf{\overline{5}}_{H}$ is in the fundamental of $SU\left(3\right)$
with a $U\left(1\right)$ charge of $-2$ so has the gauge transformation
\begin{equation}
\delta\varphi_{\mathbf{\overline{5}}_{H}}=\left(\begin{array}{ccc}
-2y & 1 & 0\\
0 & -2y & 1\\
x & 0 & -2y\end{array}\right)\left(\begin{array}{c}
a\\
b\\
c\end{array}\right),\end{equation}
 giving a matter curve of $f=x-8y^{3}$, and we can here set to zero
the first two components of the triplet $\varphi_{\mathbf{\overline{5}}_{H}}$,
and we have \begin{equation}
\eta_{\mathbf{\overline{5}}_{H}}=\left(\begin{array}{ccc}
4y^{2} & 2y & 1\\
x & 4y^{2} & 2y\\
2xy & x & 4y^{2}\end{array}\right)\left(\begin{array}{c}
0\\
0\\
\varphi_{\mathbf{\overline{5}}_{H}}^{3}\end{array}\right)=\left(\begin{array}{c}
\varphi_{\mathbf{\overline{5}}_{H}}^{3}\\
2y\varphi_{\mathbf{\overline{5}}_{H}}^{3}\\
4y^{2}\varphi_{\mathbf{\overline{5}}_{H}}^{3}\end{array}\right).\end{equation}
 The $\mathbf{5}_{H}\cdot\mathbf{10}_{M}\cdot\mathbf{10}_{M}$ Yukawa
is then \begin{equation}
W_{\mathbf{5}_{H}\cdot\mathbf{10}_{M}\cdot\mathbf{10}_{M}}=\mathrm{\tmop{Res}}_{\left(0,0\right)}\left[\frac{\mathrm{\tmop{Tr}}\left(\left[\eta_{\mathbf{5}_{H}},\eta_{\mathbf{10}_{M}}\right]\varphi_{\mathbf{10}_{M}}\right)}{\left(8y^{3}+x\right)\left(64y^{3}-x\right)}\right].\end{equation}
 Using the fact that the trace in the adjoint of $\mathfrak{e}_{7}$
(using $i,j,k$ for $SU\left(5\right)$ indices and $a,b,c$ for $SU\left(3\right)$
indices) is \begin{equation}
\mathrm{\tmop{Tr}}\left(\left[t_{\mathbf{5}i}^{a},t_{\mathbf{10}jk}^{b}\right]t_{\mathbf{10}lm}^{c}\right)\propto\epsilon_{ijklm}\epsilon^{abc}\end{equation}
 and so the Yukawa becomes \begin{equation}
W_{\mathbf{5}_{H}\cdot\mathbf{10}_{M}\cdot\mathbf{10}_{M}}=\mathrm{\tmop{Res}}_{\left(0,0\right)}\left[\frac{\epsilon_{ijklm}y\varphi_{\mathbf{5}_{H}}^{1i}\varphi_{\mathbf{10}_{M}}^{1jk}\varphi_{\mathbf{10}_{M}}^{1lm}}{\left(8y^{3}+x\right)\left(64y^{3}-x\right)}\right],\end{equation}
 which is zero as it is holomorphic in $y$ at $y=0$. Note that in
general any Yukawa coupling where all the matter curves are of the
form \begin{equation}
f=ay^{n}+bx\label{bad matter curve}\end{equation}
 with $n\geq2$ and $a,b$ are arbitrary constants, will always be
zero since the modes are always holomorphic and there will be no singularity
at $y=0$. This can be seen by noting that we can always move closer
to the origin and of course therefore we can always choose $\left|ay^{n}\right|<\left|bx\right|$
and so we can factor out the $bx$ and taylor expand to get a power
series in $y^{n}$ which is clearly holomorphic in $y$ so giving
a zero residue.

The other required Yukawa - $\mathbf{\overline{5}}_{H}\cdot\mathbf{\overline{5}}_{M}\cdot\mathbf{10}_{M}$,
works out to be \begin{equation}
W_{\mathbf{\overline{5}}_{H}\cdot\mathbf{\overline{5}}_{M}\cdot\mathbf{10}_{M}}=\mathrm{\tmop{Res}}_{\left(0,0\right)}\left[\frac{y^{2}\varphi_{\mathbf{\overline{5}}_{H}}^{3}\varphi_{\mathbf{\overline{5}}_{M}}\varphi_{\mathbf{10}_{M}}^{1}}{\left(y\right)\left(x-8y^{3}\right)}\right]=\mathrm{\tmop{Res}}_{\left(0,0\right)}\left[\frac{y\varphi_{\mathbf{\overline{5}}_{H}}^{3}\varphi_{\mathbf{\overline{5}}_{M}}\varphi_{\mathbf{10}_{M}}^{1}}{\left(x-8y^{3}\right)}\right]=0.\end{equation}
 We note that this particular $SU\left(5\right)$ GUT derived via
breaking of the $E_{7}$ gauge group is not viable as neither of the
required Yukawa couplings is present. This result is somewhat unexpected
as it seems strange that the Yukawa couplings would vanish given that
the symmetries allow for them and the matter curves intersect as required,
we note that this is just one possible way of embedding $SU\left(5\right)$
into $E_{7}$.

We now turn our attention to the breaking via $E_{8}$. 

\subsubsection{$E_{8}\rightarrow SU\left(5\right)\times SU\left(4\right)\times U\left(1\right)$}

Consider an $SU\left(4\right)\times U\left(1\right)$ Higgs field
which preserves an unbroken $SU\left(5\right)$. The Higgs is \begin{equation}
\Phi=\left(\begin{array}{cccc}
0 & 1 & 0 & 0\\
0 & 0 & 1 & 0\\
0 & 0 & 0 & 1\\
x & 0 & 0 & 0\end{array}\right)\oplus\left(y\right).\end{equation}
 The adjoint of $E_{8}$ decomposes as \begin{eqnarray}
\mathbf{248} & \rightarrow & \left(\mathbf{24},\mathbf{1}\right)_{0}\oplus\left(\mathbf{1},\mathbf{1}\right)_{0}\oplus\left(\mathbf{1},\mathbf{4}\right)_{-5}\oplus\left(\mathbf{1},\mathbf{\overline{4}}\right)_{5}\oplus\left(\mathbf{1},\mathbf{15}\right)_{0}\oplus\left(\mathbf{\overline{5}},\mathbf{4}\right)_{3}\oplus\left(\mathbf{\overline{5}},\mathbf{6}\right)_{-2}\oplus\left(\mathbf{10},\mathbf{1}\right)_{4}\oplus\left(\mathbf{10},\mathbf{4}\right)_{-1}\nonumber \\
 &  & \oplus\left(\mathbf{5},\mathbf{\overline{4}}\right)_{-3}\oplus\left(\mathbf{5},\mathbf{6}\right)_{2}\oplus\left(\mathbf{\overline{10}},\mathbf{1}\right)_{-4}\oplus\left(\mathbf{\overline{10}},\mathbf{\overline{4}}\right)_{1}\end{eqnarray}
 Looking at the $U\left(1\right)$ charges and the required couplings
we can identify the $\left(\mathbf{10},\mathbf{4}\right)_{-1}$ as
the $\mathbf{10}_{M}$, the $\left(\mathbf{5},\mathbf{6}\right)_{2}$
as the $\mathbf{5}_{H}$, the $\left(\mathbf{\overline{5}},\mathbf{6}\right)_{-2}$
as the $\mathbf{\overline{5}}_{H}$ and the $\left(\mathbf{\overline{5}},\mathbf{4}\right)_{3}$
as the $\mathbf{\overline{5}}_{M}$.

Writing the $\mathbf{10}_{M}$ mode as \begin{equation}
\varphi_{\mathbf{10}_{M}}=\left(\begin{array}{c}
\varphi_{\mathbf{10}_{M}}^{1}\\
\varphi_{\mathbf{10}_{M}}^{2}\\
\varphi_{\mathbf{10}_{M}}^{3}\\
\varphi_{\mathbf{10}_{M}}^{4}\end{array}\right),\end{equation}
 an $SU\left(4\right)$ gauge transformation acts on it as \begin{equation}
\delta\varphi_{\mathbf{10}_{M}}=\left(\begin{array}{cccc}
-y & 1 & 0 & 0\\
0 & -y & 1 & 0\\
0 & 0 & -y & 1\\
x & 0 & 0 & -y\end{array}\right)\left(\begin{array}{c}
a\\
b\\
c\\
d\end{array}\right).\end{equation}
 We can choose all but the last component of $\varphi_{\mathbf{10}_{M}}$
to be zero on the matter curve $y^{4}-x=0$. As with the $E_{7}$
model, the $\mathbf{10}_{M}$ curve is self-intersecting as required.

We can multiply by the adjugate matrix to solve the torsion equation
and obtain the localised mode: \begin{equation}
\eta_{\mathbf{10}_{M}}=\left(\begin{array}{cccc}
-y^{3} & -y^{2} & -y & -1\\
-x & -y^{3} & -y^{2} & -y\\
-xy & -x & -y^{3} & -y^{2}\\
-xy^{2} & -xy & -x & -y^{3}\end{array}\right)\left(\begin{array}{c}
0\\
0\\
0\\
\varphi_{\mathbf{10}_{M}}^{4}\end{array}\right)=\left(\begin{array}{c}
-\varphi_{\mathbf{10}_{M}}^{4}\\
-y\varphi_{\mathbf{10}_{M}}^{4}\\
-y^{2}\varphi_{\mathbf{10}_{M}}^{4}\\
-y^{3}\varphi_{\mathbf{10}_{M}}^{4}\end{array}\right).\end{equation}
 The $\mathbf{\overline{5}}_{M}$ is also in the fundamental of the
$SU\left(4\right)$ and transforms just as the $\mathbf{10}_{M}$
but with a different $U\left(1\right)$ charge, so one just replaces
$y$ with $-3y$ in the above result, yielding \begin{equation}
\eta_{\mathbf{\overline{5}}_{M}}=\left(\begin{array}{c}
-\varphi_{\mathbf{\overline{5}}_{M}}^{4}\\
3y\varphi_{\mathbf{\overline{5}}_{M}}^{4}\\
-9y^{2}\varphi_{\mathbf{\overline{5}}_{M}}^{4}\\
27y^{3}\varphi_{\mathbf{\overline{5}}_{M}}^{4}\end{array}\right),\end{equation}
 with a matter curve $f=81y^{4}-x$.

The $\mathbf{5}_{H}$, however, transforms in the $\mathbf{6}$ of
$SU\left(4\right)$ and we write it in components as \begin{equation}
\varphi_{\mathbf{5}_{H}}=\varphi_{\mathbf{5}_{H}}^{ab}e_{a}\wedge e_{b},\end{equation}
 where similar to the $SU\left(3\right)$ case above, $e_{a}$ span
the fundamental of $SU\left(4\right)$ and the basis elements $e_{a}\wedge e_{b}$
each transform under $\Phi$ as \begin{equation}
\Phi\left(e_{a}\wedge e_{b}\right)=\left(\Phi e_{a}\right)\wedge e_{b}+e_{a}\wedge\left(\Phi e_{b}\right).\end{equation}
 We write the elements of $\varphi_{\mathbf{5}_{H}}$ as a six dimensional
vector with the following basis: \begin{equation}
\left(\begin{array}{c}
e_{1}\wedge e_{2}\\
e_{1}\wedge e_{3}\\
e_{1}\wedge e_{4}\\
e_{2}\wedge e_{3}\\
e_{2}\wedge e_{4}\\
e_{3}\wedge e_{4}\end{array}\right).\end{equation}
 Under a gauge transformation, $\varphi_{\mathbf{5}_{H}}$ transforms
as \begin{equation}
\delta\varphi_{\mathbf{5}_{H}}=\left(\begin{array}{cccccc}
4y & 1 & 0 & 0 & 0 & 0\\
0 & 4y & 1 & 1 & 0 & 0\\
0 & 0 & 4y & 0 & 1 & 0\\
0 & 0 & 0 & 4y & 1 & 0\\
-x & 0 & 0 & 0 & 4y & 1\\
0 & -x & 0 & 0 & 0 & 4y\end{array}\right)\left(\begin{array}{c}
a\\
b\\
c\\
d\\
e\\
f\end{array}\right)=\left(\begin{array}{c}
4ya+b\\
4yb+c+d\\
4yc+e\\
4yd+e\\
-xa+4ye+f\\
-xb+4yf\end{array}\right),\end{equation}
 from which we can see that the matter curve has a factorisable form
$64y^{2}\left(64y^{4}+x\right)=0$.

We also see that we can use this gauge transformation to set the 1st,
2nd, 5th and one of the 3rd or 4th components to zero, we choose to
set the third to zero. Using the adjugate matrix we solve the torsion
equation as \begin{eqnarray}
\nonumber \\\eta_{\mathbf{5}_{H}} & = & \left(\begin{array}{cccccc}
8xy+1024y^{5} & -256y^{4} & 64y^{3} & 64y^{3} & -32y^{2} & 8y\\
32xy^{2} & 1024y^{5} & -256y^{4} & -256y^{4} & 128y^{3} & -32y^{2}\\
-64xy^{3} & 32xy^{2} & 8xy+1024y^{5} & -8xy & -256y^{4} & 64y^{3}\\
-64xy^{3} & 32xy^{2} & -8xy & 8xy+1024y^{5} & -256y^{4} & 64y^{3}\\
256xy^{4} & -128xy^{3} & 32xy^{2} & 32xy^{2} & 1024y^{5} & -256y^{4}\\
8x^{2}y & 256xy^{4} & -64xy^{3} & -64xy^{3} & 32xy^{2} & 8xy+1024y^{5}\end{array}\right)\left(\begin{array}{c}
0\\
0\\
0\\
\varphi_{\mathbf{5}_{H}}^{23}\\
0\\
\varphi_{\mathbf{5}_{H}}^{34}\end{array}\right)\nonumber \\
 & = & \left(\begin{array}{c}
64y^{3}\varphi_{\mathbf{5}_{H}}^{23}+8y\varphi_{\mathbf{5}_{H}}^{34}\\
-256y^{4}\varphi_{\mathbf{5}_{H}}^{23}-32y^{2}\varphi_{\mathbf{5}_{H}}^{34}\\
-8xy\varphi_{\mathbf{5}_{H}}^{23}+64y^{3}\varphi_{\mathbf{5}_{H}}^{34}\\
\left(8xy+1024y^{5}\right)\varphi_{\mathbf{5}_{H}}^{23}+64y^{3}\varphi_{\mathbf{5}_{H}}^{34}\\
32xy^{2}\varphi_{\mathbf{5}_{H}}^{23}-256y^{4}\varphi_{\mathbf{5}_{H}}^{34}\\
-64xy^{3}\varphi_{\mathbf{5}_{H}}^{23}+\left(8xy+1024y^{5}\right)\varphi_{\mathbf{5}_{H}}^{34}\end{array}\right).\end{eqnarray}
 The $\mathbf{\overline{5}}_{H}$ transforms the same as the $\mathbf{5}_{H}$
under $SU\left(4\right)$, but with the opposite $U\left(1\right)$
charge, so it has the same matter curve of $f=64y^{2}\left(64y^{4}+x\right)$
and the localized mode is given by \begin{equation}
\eta_{\mathbf{\overline{5}}_{H}}=\left(\begin{array}{c}
-64y^{3}\varphi_{\mathbf{\overline{5}}_{H}}^{23}-8y\varphi_{\mathbf{\overline{5}}_{H}}^{34}\\
-256y^{4}\varphi_{\mathbf{\overline{5}}_{H}}^{23}-32y^{2}\varphi_{\mathbf{\overline{5}}_{H}}^{34}\\
8xy\varphi_{\mathbf{\overline{5}}_{H}}^{23}-64y^{3}\varphi_{\mathbf{\overline{5}}_{H}}^{34}\\
-\left(8xy+1024y^{5}\right)\varphi_{\mathbf{\overline{5}}_{H}}^{23}-64y^{3}\varphi_{\mathbf{\overline{5}}_{H}}^{34}\\
32xy^{2}\varphi_{\mathbf{\overline{5}}_{H}}^{23}-256y^{4}\varphi_{\mathbf{\overline{5}}_{H}}^{34}\\
64xy^{3}\varphi_{\mathbf{\overline{5}}_{H}}^{23}-\left(8xy+1024y^{5}\right)\varphi_{\mathbf{\overline{5}}_{H}}^{34}\end{array}\right).\end{equation}
 The $\mathbf{5}_{H}\cdot\mathbf{10}_{M}\cdot\mathbf{10}_{M}$ Yukawa
is then \begin{equation}
W_{\mathbf{5}_{H}\cdot\mathbf{10}_{M}\cdot\mathbf{10}_{M}}=\mathrm{\tmop{Res}}_{\left(0,0\right)}\left[\frac{\mathrm{\tmop{Tr}}\left(\left[\eta_{\mathbf{5}_{H}},\eta_{\mathbf{10}_{M}}\right]\varphi_{\mathbf{10}_{M}}\right)}{y^{2}\left(64y^{4}+x\right)\left(y^{4}-x\right)}\right].\end{equation}
 The trace in $\mathfrak{e}_{8}$, using $i,j,k$ for $SU\left(5\right)$
indices and $a,b,c$ for $SU\left(4\right)$ indices is \begin{equation}
\mathrm{\tmop{Tr}}\left(\left[t_{\mathbf{5}i}^{ab},t_{\mathbf{10}jk}^{c}\right]t_{\mathbf{10}_{lm}}^{d}\right)\propto\epsilon_{ijklm}\epsilon^{abcd}.\end{equation}
 So then the Yukawa becomes \begin{eqnarray}
W_{\mathbf{5}_{H}\cdot\mathbf{10}_{M}\cdot\mathbf{10}_{M}} & = & \mathrm{\tmop{Res}}_{\left(0,0\right)}\left[\frac{\epsilon_{ijklm}\left(64y^{3}\varphi_{\mathbf{5}_{H}}^{23i}+8y\varphi_{\mathbf{5}_{H}}^{34i}\right)y^{2}\varphi_{\mathbf{10}_{M}}^{4jk}\varphi_{\mathbf{10}_{M}}^{4lm}}{\left(y^{2}\left(64y^{4}+x\right)\right)\left(y^{4}-x\right)}\right]\nonumber \\
 &  & +\mathrm{\tmop{Res}}_{\left(0,0\right)}\left[\frac{\epsilon_{ijklm}\left(256y^{4}\varphi_{\mathbf{5}_{H}}^{23i}+32y^{2}\varphi_{\mathbf{5}_{H}}^{34i}\right)y\varphi_{\mathbf{10}_{M}}^{4jk}\varphi_{\mathbf{10}_{_{M}}}^{4lm}}{\left(y^{2}\left(64y^{4}+x\right)\right)\left(y^{4}-x\right)}\right]\nonumber \\
 &  & +\mathrm{\tmop{Res}}_{\left(0,0\right)}\left[\frac{\epsilon_{ijklm}\left(\left(8xy+1024y^{5}\right)\varphi_{\mathbf{5}_{H}}^{23i}+64y^{3}\varphi_{\mathbf{5}_{H}}^{34i}\right)\varphi_{\mathbf{10}_{M}}^{4jk}\varphi_{\mathbf{10}_{M}}^{4lm}}{\left(y^{2}\left(64y^{4}+x\right)\right)\left(y^{4}-x\right)}\right].\end{eqnarray}
 Which, after simplifying, becomes \begin{equation}
W_{\mathbf{5}_{H}\cdot\mathbf{10}_{M}\cdot\mathbf{10}_{M}}=\mathrm{\tmop{Res}}_{\left(0,0\right)}\left[\frac{\epsilon_{ijklm}\varphi_{\mathbf{5}_{H}}^{23i}\varphi_{\mathbf{10}_{M}}^{4jk}\varphi_{\mathbf{10}_{M}}^{4lm}}{\left(x\right)\left(y\right)}\right].\end{equation}
 Using the trace result \begin{equation}
\mathrm{\tmop{Tr}}\left(\left[t_{\mathbf{\overline{5}}i}^{a},t_{\mathbf{10}jk}^{b}\right]t_{\mathbf{\overline{5}}l}^{cd}\right)\propto\delta_{ij}\delta_{kl}\epsilon^{abcd},\end{equation}
 the $\mathbf{\overline{5}}_{H}\cdot\mathbf{\overline{5}}_{M}\cdot\mathbf{10}_{M}$
Yukawa works out to be \begin{equation}
W_{\mathbf{\overline{5}}_{H}\cdot\mathbf{\overline{5}}_{M}\cdot\mathbf{10}_{M}}=\mathrm{\tmop{Res}}_{\left(0,0\right)}\left[\frac{\varphi_{\mathbf{\overline{5}}_{H}i}^{23}\varphi_{\mathbf{\overline{5}}_{M}j}^{4}\varphi_{\mathbf{10}_{M}}^{4ij}}{\left(x\right)\left(y\right)}\right].\end{equation}

As advertised in the Introduction, one important feature for the models
with $E_{7}$ or $E_{8}$ breaking is the absence of the proton decay
terms. For the $E_{8}$ case discussed in this subsection, the 4-dimensional
proton decay mediating operators would be of the form $\mathbf{\overline{5}}_{M}\cdot\mathbf{10}_{M}\cdot\mathbf{\overline{5}}_{M}$
or $\mathbf{5}_{M}\cdot\mathbf{10}_{M}\cdot\mathbf{10}_{M}$. Without
any further computations, we see from the definitions of the field
charges that these terms are forbidden because they do have the allowed
$U\left(1\right)$ charges, as discussed in \cite{Marsano:2009gv}.

\subsection{Right-handed Neutrinos }

Another important coupling is that of the right-handed neutrino, which
is a singlet of $SU\left(5\right)$. It can couple in either Dirac
or Majorana way. The Dirac scenario requires just one coupling of
the form $\mathbf{5}_{H}\cdot\mathbf{\overline{5}}_{M}\cdot\mathbf{1}$.
As considered in reference {[}5{]}, the right handed neutrinos can
be seen as complex structure deformations. If F-theory is compactified
on $X$, a Calabi-Yau 4-fold, the $SU(5)_{GUT}$ singlet field in
the singlet Yukawa coupling was considered to be related to fluctuations
from the vacuum in $H^{1,2}(X)$ and the Yukawa coupling was calculated
by an overlap integration \begin{equation}
\int_{S}\mbox{tr}(\chi_{\mathbf{6}}\psi_{\mathbf{15}}\psi_{\mathbf{4}})\end{equation}
 where $\mathbf{6},\mathbf{15},\mathbf{4}$ refer to the representations
of the transverse $SU(4)$. If we identify the right handed neutrinos
with the adjoint representation of the transverse group, then we cannot
use the residue formula to compute its Yukawa coupling. This is expected
as the right-handed neutrino is not localized on matter curve but
corresponds to deformations of the complex structure.

On the other hand, the results of {[}5{]} were based on the fact that
the field $\Phi$ is diagonal when the deformation of the complex
structure of $X$ of an F-theory compactification correspond to the
the (2,0) forms in the Cartan part of the transverse group. By using
the T-brane formalism, this consideration should be changed and one
needs to rethink the issue of identifying the complex structure deformations.
We leave this issue for a future publication.


\section{Conclusions}

In the present work we have presented some models of T-branes which
correspond to brane configurations with $Z_{3}$ and $Z_{4}$ monodromies.
These configurations have been used to break the $E_{7}$ and $E_{8}$
groups to the $SU(5)$ grand unification group. We used the residue
formulas to compute Yukawa couplings for both $E_{7}\rightarrow SU(5)\times SU(3)\times U(1)$
and $E_{8}\rightarrow SU(5)\times SU(4)\times U(1)$.

There are two interesting directions which can be followed. The first
one involves obtaining a solution to the differential equation for
$Z_{3}$ and $Z_{4}$ background derived in our work. They should
be generalizations of the Painleve III differential equations and
would allow an explicit solution for the supersymmetric brane configurations.
The second direction is to obtain an understanding of right handed
neutrinos in the context of T-branes where one goes beyond the Cartan
subalgebra. This would allow further insights into Yukawa couplings
for right handed neutrinos.

This is a particularly relevant issue fo F-theory studies, due to
the absence of adjoint and higher order scalar representations in
perturbative heterotic constructions \cite{Dienes:1996yh,Dienes:1996wx},
specifically, the absence of the $\mathbf{126}$ representation of
$SO\left(10\right)$ in perturbative constructions indicates that
the right-handed neutrino Majorana mass can only be generated by a
vev of a Higgs field in the $\mathbf{16}$ representation, hence breaking
lepton number by one unit. The interesting question therefore is whether
the non-perturbative framework of F-theory offers some new possibilities.

\section*{Acknowledgements}


We would like to thank Taizan Watari for important discussions over
related subjects. This work is supported by the STFC under contract
PP/D000416/1. We would also like to thank our referee for helpful
suggestions.



\appendix

\section{Appendix}

\subsection{Details of $SU\left(3\right)$ Computations}

We present here the details of the computations for the SU(3) case
corresponding to a $Z_{3}$ T-brane. Firstly we work out our new Higgs
field in unitary gauge: \begin{equation}
\Phi=g\left(\begin{array}{ccc}
0 & 1 & 0\\
0 & 0 & 1\\
0 & 0 & 0\end{array}\right)g^{-1}=\left(\begin{array}{ccc}
0 & e^{f_{1}-f_{2}} & 0\\
0 & 0 & e^{f_{2}-f_{3}}\\
0 & 0 & 0\end{array}\right).\end{equation}
 then the commutator part of the D-term equation which is \begin{equation}
\left[\Phi^{\dagger},\Phi\right]=\left(\begin{array}{ccc}
-e^{2\left(f_{1}-f_{2}\right)} & 0 & 0\\
0 & e^{2\left(f_{1}-f_{2}\right)}-e^{2\left(f_{2}-f_{3}\right)} & 0\\
0 & 0 & e^{2\left(f_{2}-f_{3}\right)}\end{array}\right),\end{equation}
 But the Toda equation requires only two independent function, we
denote them by $e^{h_{1}}$and $e^{h_{2}}$, with \begin{eqnarray}
h_{1} & = & 2\left(f_{1}-f_{2}\right)\nonumber \\
h_{2} & = & 2\left(f_{2}-f_{3}\right).\end{eqnarray}
 Combined with $f_{1}+f_{2}+f_{3}=0$, we obtain \begin{eqnarray}
f_{1} & = & \frac{1}{6}\left(2h_{1}+h_{2}\right)\nonumber \\
f_{2} & = & \frac{1}{6}\left(h_{2}-h_{1}\right)\nonumber \\
f_{3} & = & \frac{1}{6}\left(-h_{1}-2h_{2}\right).\end{eqnarray}
 Hence the required unitary transformation, reverting to $f_{a}$
instead of $h_{a}$ for the linearly independent functions, is \begin{equation}
g=\left(\begin{array}{ccc}
e^{\frac{1}{6}\left(2f_{1}+f_{2}\right)} & 0 & 0\\
0 & e^{\frac{1}{6}\left(f_{2}-f_{1}\right)} & 0\\
0 & 0 & e^{\frac{1}{6}\left(-f_{1}-2f_{2}\right)}\end{array}\right).\end{equation}
 This gives a transformed Higgs field \begin{equation}
\Phi=\left(\begin{array}{ccc}
0 & e^{\frac{1}{2}f_{1}} & 0\\
0 & 0 & e^{\frac{1}{2}f_{2}}\\
0 & 0 & 0\end{array}\right),\end{equation}
 and the commutator part of the D-term equation \begin{equation}
\left[\Phi^{\dagger},\Phi\right]=\left(\begin{array}{ccc}
-e^{f_{1}} & 0 & 0\\
0 & e^{f_{1}}-e^{f_{2}} & 0\\
0 & 0 & e^{f_{2}}\end{array}\right).\end{equation}
 The connection is \begin{equation}
A^{0,1}=g\overline{\partial}g^{-1}=\frac{1}{6}\left(\begin{array}{ccc}
-2\overline{\partial}f_{1}-\overline{\partial}f_{2} & 0 & 0\\
0 & \overline{\partial}f_{1}-\overline{\partial}f_{2} & 0\\
0 & 0 & \overline{\partial}f_{1}+2\overline{\partial}f_{2}\end{array}\right),\end{equation}
 and the D-term equations become \begin{eqnarray}
\frac{1}{3}\left(-2\partial\overline{\partial}f_{1}-\partial\overline{\partial}f_{2}\right) & = & -e^{f_{1}}\nonumber \\
\frac{1}{3}\left(\partial\overline{\partial}f_{1}-\partial\overline{\partial}f_{2}\right) & = & e^{f_{1}}-e^{f_{2}}\nonumber \\
\frac{1}{3}\left(\partial\overline{\partial}f_{1}+2\partial\overline{\partial}f_{2}\right) & = & e^{f_{2}}.\end{eqnarray}
 Notice that there are actually just two equations as the third is
just the negative of the sum of the other two.

One can then take linear combinations of these equations to obtain
\begin{eqnarray}
\partial\overline{\partial}f_{1} & = & 2e^{f_{1}}-e^{f_{2}}\nonumber \\
\partial\overline{\partial}f_{2} & = & -e^{f_{1}}+2e^{f_{2}}.\end{eqnarray}

\subsubsection{General case}

For the more general case of \begin{equation}
\Phi=\left(\begin{array}{ccc}
0 & 1 & 0\\
0 & 0 & 1\\
x & 0 & 0\end{array}\right),\end{equation}
 we need to take a more complicated unitary transformation \begin{equation}
g=\left(\begin{array}{ccc}
r^{m}e^{\frac{1}{6}\left(2f_{1}+f_{2}\right)} & 0 & 0\\
0 & r^{n}e^{\frac{1}{6}\left(f_{2}-f_{1}\right)} & 0\\
0 & 0 & r^{-m-n}e^{\frac{1}{6}\left(-f_{1}-2f_{2}\right)}\end{array}\right),\end{equation}
 where $x=re^{i\theta}$ and the numbers $m$ and $n$ are determined
by demanding that the second term in the D-term equation be homogeneous
in $r$. The $f_{a}$ are assumed to be independent of $y$ and $\theta$.

The transformed Higgs field is \begin{equation}
\Phi=g\left(\begin{array}{ccc}
0 & 1 & 0\\
0 & 0 & 1\\
x & 0 & 0\end{array}\right)g^{-1}=\left(\begin{array}{ccc}
0 & r^{m-n}e^{\frac{1}{2}f_{1}} & 0\\
0 & 0 & r^{m+2n}e^{\frac{1}{2}f_{2}}\\
xr^{-2m-n}e^{\frac{1}{2}\left(-f_{1}-f_{2}\right)} & 0 & 0\end{array}\right).\end{equation}
 This gives \begin{equation}
\left[\Phi^{\dagger},\Phi\right]=\left(\begin{array}{ccc}
r^{2-4m-2n}e^{-f_{1}-f_{2}}-r^{2m-2n}e^{f_{1}} & 0 & 0\\
0 & r^{2m-2n}e^{f_{1}}-r^{2m+4n}e^{f_{2}} & 0\\
0 & 0 & r^{2m+4n}e^{f_{2}}-r^{2-4m-2n}e^{-f_{1}-f_{2}}\end{array}\right),\end{equation}
 and homogeneity in $r$ requires \begin{eqnarray}
2-4m-2n & = & 2m-2n\nonumber \\
2m-2n & = & 2m+4n,\end{eqnarray}
 which implies $m=\frac{1}{3},\hspace{0.25em}n=0,$ and therefore
the unitary transformation is \begin{equation}
g=\left(\begin{array}{ccc}
r^{\frac{1}{3}}e^{\frac{1}{6}\left(2f_{1}+f_{2}\right)} & 0 & 0\\
0 & e^{\frac{1}{6}\left(f_{2}-f_{1}\right)} & 0\\
0 & 0 & r^{-\frac{1}{3}}e^{\frac{1}{6}\left(-f_{1}-2f_{2}\right)}\end{array}\right),\end{equation}
 and so \begin{equation}
\left[\Phi^{\dagger},\Phi\right]=\left(\begin{array}{ccc}
r^{\frac{2}{3}}e^{-f_{1}-f_{2}}-r^{\frac{2}{3}}e^{f_{1}} & 0 & 0\\
0 & r^{\frac{2}{3}}e^{f_{1}}-r^{\frac{2}{3}}e^{f_{2}} & 0\\
0 & 0 & r^{\frac{2}{3}}e^{f_{2}}-r^{\frac{2}{3}}e^{-f_{1}-f_{2}}\end{array}\right).\end{equation}
 Then, using \begin{eqnarray}
\frac{\partial}{\partial\overline{x}} & = & e^{i\theta}\frac{\partial}{\partial r}+\frac{ie^{i\theta}}{r}\frac{\partial}{\partial\theta}\nonumber \\
\frac{\partial}{\partial x} & = & e^{-i\theta}\frac{\partial}{\partial r}-\frac{ie^{-i\theta}}{r}\frac{\partial}{\partial\theta},\end{eqnarray}
 we derive the connection to be \begin{equation}
A^{0,1}=g\overline{\partial}g^{-1}=\left(\begin{array}{ccc}
-\frac{e^{i\theta}}{3r}-\frac{1}{6}\left(2\overline{\partial}f_{1}+\overline{\partial}f_{2}\right) & 0 & 0\\
0 & \frac{1}{6}\left(\overline{\partial}f_{1}-\overline{\partial}f_{2}\right) & 0\\
0 & 0 & \frac{e^{i\theta}}{3r}+\frac{1}{6}\left(\overline{\partial}f_{1}+2\overline{\partial}f_{2}\right)\end{array}\right),\end{equation}
 and also \begin{equation}
F_{A}^{1,1}=\left(\begin{array}{ccc}
-\frac{1}{3}\left(2\partial\overline{\partial}f_{1}+\partial\overline{\partial}f_{2}\right) & 0 & 0\\
0 & \frac{1}{3}\left(\partial\overline{\partial}f_{1}-\partial\overline{\partial}f_{2}\right) & 0\\
0 & 0 & \frac{1}{3}\left(\partial\overline{\partial}f_{1}+2\partial\overline{\partial}f_{2}\right)\end{array}\right).\end{equation}
 This means that the D-term equation gives \begin{eqnarray}
\frac{1}{3}\left(-2\partial\overline{\partial}f_{1}-\partial\overline{\partial}f_{2}\right) & = & r^{\frac{2}{3}}\left(e^{f_{1}-f_{2}}-e^{f_{1}}\right)\nonumber \\
\frac{1}{3}\left(\partial\overline{\partial}f_{1}-\partial\overline{\partial}f_{2}\right) & = & r^{\frac{2}{3}}\left(e^{f_{1}}-e^{f_{2}}\right)\nonumber \\
\frac{1}{3}\left(\partial\overline{\partial}f_{1}+2\partial\overline{\partial}f_{2}\right) & = & r^{\frac{2}{3}}\left(e^{f_{2}}-e^{-f_{1}-f_{2}}\right).\end{eqnarray}
 Again these three equations are just two independent ones, which
we get after taking linear combinations and using the fact that we
defined the $f_{a}$ to only depend on $r$ as: \begin{eqnarray}
\left(\frac{d^{2}}{dr^{2}}+\frac{1}{r}\frac{d}{dr}\right)f_{1} & = & r^{\frac{2}{3}}\left(2e^{f_{1}}-e^{-f_{1}-f_{2}}-e^{f_{2}}\right)\nonumber \\
\left(\frac{d^{2}}{dr^{2}}+\frac{1}{r}\frac{d}{dr}\right)f_{2} & = & r^{\frac{2}{3}}\left(2e^{f_{2}}-e^{-f_{1}-f_{2}}-e^{f_{1}}\right).\end{eqnarray}

\subsection{Details of $SU\left(4\right)$ Computations}

Let us consider the spectral equation for an $SU(4)$ field: \begin{equation}
P_{\phi}(z)=z^{4}-x,\end{equation}
 for which there is a $Z_{4}$ monodromy. In the holomorphic gauge
the Higgs field becomes \begin{equation}
\Phi=\left(\begin{array}{cccc}
0 & 1 & 0 & 0\\
0 & 0 & 1 & 0\\
0 & 0 & 0 & 1\\
x & 0 & 0 & 0\end{array}\right).\end{equation}
 which is an intermediate between a diagonal Higgs field and a nilpotent
Higgs field in holomorphic gauge \begin{equation}
\Phi=\left(\begin{array}{cccc}
0 & 1 & 0 & 0\\
0 & 0 & 1 & 0\\
0 & 0 & 0 & 1\\
0 & 0 & 0 & 0\end{array}\right),\end{equation}
 The transition to the unitary gauge is achieved by using \begin{equation}
g=\left(\begin{array}{cccc}
e^{f_{1}} & 0 & 0 & 0\\
0 & e^{f_{2}} & 0 & 0\\
0 & 0 & e^{f_{3}} & 0\\
0 & 0 & 0 & e^{f_{4}}\end{array}\right),\end{equation}
 with the unit determinant condition that $\sum_{a}f_{a}=0$. This
gives the Higgs field in unitary gauge as \begin{equation}
\Phi=g\left(\begin{array}{cccc}
0 & 1 & 0 & 0\\
0 & 0 & 1 & 0\\
0 & 0 & 0 & 1\\
0 & 0 & 0 & 0\end{array}\right)g^{-1}=\left(\begin{array}{cccc}
0 & e^{f_{1}-f_{2}} & 0 & 0\\
0 & 0 & e^{f_{2}-f_{3}} & 0\\
0 & 0 & 0 & e^{f_{3}-f_{4}}\\
0 & 0 & 0 & 0\end{array}\right),\end{equation}
 the commutator part of the D-term equation is then \begin{equation}
\left[\Phi^{\dagger},\Phi\right]=\left(\begin{array}{cccc}
-e^{2\left(f_{1}-f_{2}\right)} & 0 & 0 & 0\\
0 & e^{2\left(f_{1}-f_{2}\right)}-e^{2\left(f_{2}-f_{3}\right)} & 0 & 0\\
0 & 0 & e^{2\left(f_{2}-f_{3}\right)}-e^{2\left(f_{3}-f_{4}\right)} & 0\\
0 & 0 & 0 & e^{2\left(f_{3}-f_{4}\right)}\end{array}\right),\end{equation}
 and to get $g$ in terms of the linearly independent functions $h_{a}$,
we require \begin{eqnarray}
2\left(f_{1}-f_{2}\right) & = & h_{1}\nonumber \\
2\left(f_{2}-f_{3}\right) & = & h_{2}\nonumber \\
2\left(f_{3}-f_{4}\right) & = & h_{3},\end{eqnarray}
 along with \begin{equation}
f_{1}+f_{2}+f_{3}+f_{4}=0,\end{equation}
 which is solved by \begin{eqnarray}
f_{1} & = & \frac{1}{8}\left(3h_{1}+2h_{2}+h_{3}\right)\nonumber \\
f_{2} & = & \frac{1}{8}\left(-h_{1}+2h_{2}+h_{3}\right)\nonumber \\
f_{3} & = & \frac{1}{8}\left(-h_{1}-2h_{2}+h_{3}\right)\nonumber \\
f_{4} & = & \frac{1}{8}\left(-h_{1}-2h_{2}-3h_{3}\right).\end{eqnarray}
 The required unitary transformation $g$, after again writing the
independent functions as $f_{a}$ instead of $h_{a}$, is \begin{equation}
g=\left(\begin{array}{cccc}
e^{\frac{1}{8}\left(3f_{1}+2f_{2}+f_{3}\right)} & 0 & 0 & 0\\
0 & e^{\frac{1}{8}\left(-f_{1}+2f_{2}+f_{3}\right)} & 0 & 0\\
0 & 0 & e^{\frac{1}{8}\left(-f_{1}-2f_{2}+f_{3}\right)} & 0\\
0 & 0 & 0 & e^{\frac{1}{8}\left(-f_{1}-2f_{2}-3f_{3}\right)}\end{array}\right).\end{equation}
 The unitary gauge Higgs field is \begin{equation}
\Phi=\left(\begin{array}{cccc}
0 & e^{\frac{1}{2}f_{1}} & 0 & 0\\
0 & 0 & e^{\frac{1}{2}f_{2}} & 0\\
0 & 0 & 0 & e^{\frac{1}{2}f_{3}}\\
0 & 0 & 0 & 0\end{array}\right),\end{equation}
 and the commutator part of the D-term equation is \begin{equation}
\left[\Phi^{\dagger},\Phi\right]=\left(\begin{array}{cccc}
-e^{f_{1}} & 0 & 0 & 0\\
0 & e^{f_{1}}-e^{f_{2}} & 0 & 0\\
0 & 0 & e^{f_{2}}-e^{f_{3}} & 0\\
0 & 0 & 0 & e^{f_{3}}\end{array}\right).\end{equation}
 The unitary connection is given by \begin{equation}
A^{0,1}=g\overline{\partial}g^{-1}=\frac{1}{8}\left(\begin{array}{cccc}
-3\overline{\partial}f_{1}-2\overline{\partial}f_{2}-\overline{\partial}f_{3} & 0 & 0 & 0\\
0 & \overline{\partial}f_{1}-2\overline{\partial}f_{2}-\overline{\partial}f_{3} & 0 & 0\\
0 & 0 & \overline{\partial}f_{1}+2\overline{\partial}f_{2}-\overline{\partial}f_{3} & 0\\
0 & 0 & 0 & \overline{\partial}f_{1}+2\overline{\partial}f_{2}+3\overline{\partial}f_{3}\end{array}\right),\end{equation}
 and therefore the D-term equation gives \begin{eqnarray}
\frac{1}{4}\left(-3\partial\overline{\partial}f_{1}-2\partial\overline{\partial}f_{2}-\partial\overline{\partial}f_{3}\right) & = & -e^{f_{1}}\nonumber \\
\frac{1}{4}\left(\partial\overline{\partial}f_{1}-2\partial\overline{\partial}f_{2}-\partial\overline{\partial}f_{3}\right) & = & e^{f_{1}}-e^{f_{2}}\nonumber \\
\frac{1}{4}\left(\partial\overline{\partial}f_{1}+2\partial\overline{\partial}f_{2}-\partial\overline{\partial}f_{3}\right) & = & e^{f_{2}}-e^{f_{3}}\nonumber \\
\frac{1}{4}\left(\partial\overline{\partial}f_{1}+2\partial\overline{\partial}f_{2}+3\partial\overline{\partial}f_{3}\right) & = & e^{f_{3}}.\end{eqnarray}
 As with $SU\left(3\right)$ the last equation is just the negative
of the sum of the first three, and taking linear combinations one
obtains \begin{eqnarray}
\partial\overline{\partial}f_{1} & = & 2e^{f_{1}}-e^{f_{2}}\nonumber \\
\partial\overline{\partial}f_{2} & = & -e^{f_{1}}+2e^{f_{2}}-e^{f_{3}}\nonumber \\
\partial\overline{\partial}f_{3} & = & -e^{f_{2}}+2e^{f_{3}}.\end{eqnarray}
 This is of the desired form \begin{equation}
\partial\overline{\partial}f_{a}=C_{ab}e^{f_{b}},\end{equation}
 where $C_{ab}$ is the $SU\left(4\right)$ Cartan matrix \begin{equation}
\left(\begin{array}{ccc}
2 & -1 & 0\\
-1 & 2 & -1\\
0 & -1 & 2\end{array}\right).\end{equation}

\subsubsection{General case}

For the general case, we proceed the same way as with $SU\left(3\right)$,
taking a more complicated unitary transformation \begin{equation}
g=\left(\begin{array}{cccc}
r^{l}e^{\frac{1}{8}\left(3f_{1}+2f_{2}+f_{3}\right)} & 0 & 0 & 0\\
0 & r^{m}e^{\frac{1}{8}\left(-f_{1}+2f_{2}+f_{3}\right)} & 0 & 0\\
0 & 0 & r^{n}e^{\frac{1}{8}\left(-f_{1}-2f_{2}+f_{3}\right)} & 0\\
0 & 0 & 0 & r^{-l-m-n}e^{\frac{1}{8}\left(-f_{1}-2f_{2}-3f_{3}\right)}\end{array}\right).\end{equation}
 Then the unitary Higgs field is \begin{equation}
\Phi=g\left(\begin{array}{cccc}
0 & 1 & 0 & 0\\
0 & 0 & 1 & 0\\
0 & 0 & 0 & 1\\
x & 0 & 0 & 0\end{array}\right)g^{-1}=\left(\begin{array}{cccc}
0 & r^{l-m}e^{\frac{1}{2}f_{1}} & 0 & 0\\
0 & 0 & r^{m-n}e^{\frac{1}{2}f_{2}} & 0\\
0 & 0 & 0 & r^{+l+m+2n}e^{\frac{1}{2}f_{3}}\\
xr^{-2l-m-n}e^{\frac{1}{2}\left(-f_{1}-f_{2}-f_{3}\right)} & 0 & 0 & 0\end{array}\right).\end{equation}
 Requiring the commutator part of the D-term equation to be homogeneous
in $r$ implies that \begin{eqnarray}
l & = & \frac{3}{8}\nonumber \\
m & = & \frac{1}{8}\nonumber \\
n & = & -\frac{1}{8},\end{eqnarray}
 which means that the required unitary transformation is \begin{equation}
g=\left(\begin{array}{cccc}
r^{\frac{3}{8}}e^{\frac{1}{8}\left(3f_{1}+2f_{2}+f_{3}\right)} & 0 & 0 & 0\\
0 & r^{\frac{1}{8}}e^{\frac{1}{8}\left(-f_{1}+2f_{2}+f_{3}\right)} & 0 & 0\\
0 & 0 & r^{-\frac{1}{8}}e^{\frac{1}{8}\left(-f_{1}-2f_{2}+f_{3}\right)} & 0\\
0 & 0 & 0 & r^{-\frac{3}{8}}e^{\frac{1}{8}\left(-f_{1}-2f_{2}-3f_{3}\right)}\end{array}\right),\end{equation}
 and \begin{equation}
\left[\Phi^{\dagger},\Phi\right]=\left(\begin{array}{cccc}
r^{\frac{1}{2}}e^{-f_{1}-f_{2}-f_{3}}-r^{\frac{1}{2}}e^{f_{1}} & 0 & 0 & 0\\
0 & r^{\frac{1}{2}}e^{f_{1}}-r^{\frac{1}{2}}e^{f_{2}} & 0 & 0\\
0 & 0 & r^{\frac{1}{2}}e^{f_{2}}-r^{\frac{1}{2}}e^{f_{3}} & 0\\
0 & 0 & 0 & r^{\frac{1}{2}}e^{f_{3}}-r^{\frac{1}{2}}e^{-f_{1}-f_{2}-f_{3}}\end{array}\right).\end{equation}
 The connection is \begin{equation}
\frac{1}{8}\left(\begin{array}{cccc}
-\frac{3e^{i\theta}}{r}-3\overline{\partial}f_{1}-2\overline{\partial}f_{2}-\overline{\partial}f_{3} & 0 & 0 & 0\\
0 & -\frac{e^{i\theta}}{r}+\overline{\partial}f_{1}-2\overline{\partial}f_{2}-\overline{\partial}f_{3} & 0 & 0\\
0 & 0 & \frac{e^{i\theta}}{r}+\overline{\partial}f_{1}+2\overline{\partial}f_{2}-\overline{\partial}f_{3} & 0\\
0 & 0 & 0 & \frac{3e^{i\theta}}{r}+\overline{\partial}f_{1}+2\overline{\partial}f_{2}+3\overline{\partial}f_{3}\end{array}\right),\end{equation}
 and so the D-term equation is \begin{eqnarray}
\frac{1}{4}\left(-3\partial\overline{\partial}f_{1}-2\partial\overline{\partial}f_{2}-\partial\overline{\partial}f_{3}\right) & = & r^{\frac{1}{2}}\left(e^{-f_{1}-f_{2}-f_{3}}-e^{f_{1}}\right)\nonumber \\
\frac{1}{4}\left(\partial\overline{\partial}f_{1}-2\partial\overline{\partial}f_{2}-\partial\overline{\partial}f_{3}\right) & = & r^{\frac{1}{2}}\left(e^{f_{1}}-e^{f_{2}}\right)\nonumber \\
\frac{1}{4}\left(\partial\overline{\partial}f_{1}+2\partial\overline{\partial}f_{2}-\partial\overline{\partial}f_{3}\right) & = & r^{\frac{1}{2}}\left(e^{f_{2}}-e^{f_{3}}\right)\nonumber \\
\frac{1}{4}\left(\partial\overline{\partial}f_{1}+2\partial\overline{\partial}f_{2}+3\partial\overline{\partial}f_{3}\right) & = & r^{\frac{1}{2}}\left(e^{f_{3}}-e^{-f_{1}-f_{2}-f_{3}}\right).\end{eqnarray}
 As with the nilpotent case, the fourth equation here is just the
negative sum of the first three. Taking linear combinations, one obtains
\begin{eqnarray}
\partial\overline{\partial}f_{1} & = & r^{\frac{1}{2}}\left(2e^{f_{1}}-e^{f_{2}}-e^{-f_{1}-f_{2}-f_{3}}\right)\nonumber \\
\partial\overline{\partial}f_{2} & = & r^{\frac{1}{2}}\left(-e^{f_{1}}+2e^{f_{2}}-e^{f_{3}}\right)\nonumber \\
\partial\overline{\partial}f_{3} & = & r^{\frac{1}{2}}\left(-e^{f_{2}}+2e^{f_{3}}-e^{-f_{1}-f_{2}-f_{3}}\right).\end{eqnarray}

\subsection{General $SU\left(n\right)$}


Here we break $SU\left(n\right)$ with an $n\times n$ matrix-valued
Higgs field \begin{equation}
\Phi=\left(\begin{array}{ccccc}
0 & 1 & 0 & \cdots & 0\\
0 & 0 & 1 & \cdots & 0\\
\vdots & \vdots & \vdots & \ddots & \vdots\\
0 & 0 & 0 & \cdots & 1\\
x & 0 & 0 & \cdots & 0\end{array}\right).\end{equation}

\subsubsection{Nilpotent case $\left(x=0\right)$}

The nilpotent Higgs field in holomorphic gauge is \begin{equation}
\Phi=\left(\begin{array}{ccccc}
0 & 1 & 0 & \cdots & 0\\
0 & 0 & 1 & \cdots & 0\\
\vdots & \vdots & \vdots & \ddots & \vdots\\
0 & 0 & 0 & \cdots & 1\\
0 & 0 & 0 & \cdots & 0\end{array}\right).\end{equation}
 To solve the D-term equation, as with $SU\left(3\right)$ and $SU\left(4\right)$,
we move to unitary gauge with a transformation \begin{equation}
g=\left(\begin{array}{cccc}
e^{f_{1}} & 0 & \cdots & 0\\
0 & e^{f_{2}} & \cdots & 0\\
\vdots & \vdots & \ddots & \vdots\\
0 & 0 & 0 & e^{f_{n}}\end{array}\right),\end{equation}
 where the unit determinant condition for $g$ implies $\sum f_{a}=0$.

The most convenient way to express $f_{a}$ in terms of the $n-1$
linearly independent $h_{a}$'s is \begin{equation}
f_{1}=\frac{1}{2n}\left(\left(n-1\right)h_{1}+\left(n-2\right)h_{2}+\cdots2h_{n-2}+h_{n-1}\right).\label{general n first f eq}\end{equation}
 With the rest of the $f_{a}$ determined by the conditions \begin{equation}
2\left(f_{a}-f_{a+1}\right)=h_{a}.\label{general n 2nd f eq}\end{equation}
 With this particular unitary transformation, the D-term equation
simply becomes \begin{equation}
\partial\overline{\partial}h_{a}=C_{ab}e^{h_{b}},\end{equation}
 where $C_{ab}$ is the Cartan matrix of $SU\left(n\right)$.

\subsubsection{General case}

For the general case, as before, a more complicated unitary transformation
is required: \begin{equation}
g=\left(\begin{array}{cccc}
r^{m_{1}}e^{f_{1}} & 0 & \cdots & 0\\
0 & r^{m_{2}}e^{f_{2}} & \cdots & 0\\
\vdots & \vdots & \ddots & \vdots\\
0 & 0 & 0 & r^{m_{n}}e^{f_{n}}\end{array}\right),\end{equation}
 with $\sum m_{a}=0$ and the $f_{a}$ defined as in equations (\ref{general n first f eq})
and (\ref{general n 2nd f eq}). The unitary Higgs field is then \begin{equation}
\Phi=\left(\begin{array}{ccccc}
0 & r^{m_{1}-m_{2}}e^{\frac{1}{2}h_{1}} & 0 & \cdots & 0\\
0 & 0 & r^{m_{2}-m_{3}}e^{\frac{1}{2}h_{2}} & \cdots & 0\\
\vdots & \vdots & \vdots & \ddots & \vdots\\
0 & 0 & 0 & \cdots & r^{m_{n-1}-m_{n}}e^{\frac{1}{2}h_{n-1}}\\
xr^{m_{n}-m_{1}}e^{-\frac{1}{2}\left(\sum h_{a}\right)} & 0 & 0 & \cdots & 0\end{array}\right),\end{equation}
 and the commutator part of the D-term equation is \begin{equation}
\left(\begin{array}{ccccc}
r^{2+2m_{n}-2m_{1}}e^{-\sum h_{a}}-r^{2m_{1}-2m_{2}}e^{h_{1}} & 0 & \cdots & 0\\
0 & r^{2m_{1}-2m_{2}}e^{h_{1}}-r^{2m_{2}-2m_{3}}e^{h_{2}} & \cdots & 0\\
\vdots & \vdots & \ddots & \vdots\\
0 & 0 & \cdots & r^{2m_{n-1}-2m_{n}}e^{h_{n-1}}-r^{2+2m_{n}-2m_{1}}e^{-\sum h_{a}}\end{array}\right).\end{equation}
 The D-term can then be made homogeneous in $r$ by choosing \begin{equation}
m_{a}=\frac{n+1-2a}{2n}.\end{equation}
 Then D-term equation then becomes \begin{eqnarray}
\left(\frac{d^{2}}{dr^{2}}+\frac{1}{r}\frac{d}{dr}\right)h_{1} & = & r^{\frac{2}{n}}\left(-e^{-\sum h_{a}}+2e^{h_{1}}-e^{h_{2}}\right)\nonumber \\
\left(\frac{d^{2}}{dr^{2}}+\frac{1}{r}\frac{d}{dr}\right)h_{a} & = & r^{\frac{2}{n}}C_{ab}e^{h_{b}}\hspace{0.25em}\hspace{0.25em}\hspace{0.25em}\hspace{0.25em}\hspace{0.25em}\hspace{0.25em}\hspace{0.25em}\hspace{0.25em}\hspace{0.25em}\hspace{0.25em}a=2,\ldots,n-2\nonumber \\
\left(\frac{d^{2}}{dr^{2}}+\frac{1}{r}\frac{d}{dr}\right)h_{n-1} & = & r^{\frac{2}{n}}\left(-e^{h_{n-2}}+2e^{h_{n-1}}-e^{-\sum h_{a}}\right).\end{eqnarray}
 \bibliographystyle{unsrt} \bibliographystyle{unsrt}
\bibliography{TbranesandYukawaCouplingsv2}

\end{document}